# Real Time scheduling with Virtual Nodes for Self Stabilization in Wireless Sensor Networks

Deepali Virmani [1] , Satbir Jain [2]

[1] Department of computer science, BPIT, IPU, Delhi, India

[2] Department of computer science, NSIT, DU, Delhi, India

deepalivirmani@gmail.com

**Abstract**

In this paper we propose a new scheduling algorithm called Real Time Scheduling (RTS) which uses virtual nodes for self stabilization. This algorithm deals with all the contributing components of the end-to-end travelling delay of data packets in sensor network and with virtual nodes algorithm achieves QoS in terms of packet delivery, multiple connections, better power management and stable routes in case of failure. RTS delays packets at intermediate hops (not just prioritizes them) for a duration that is a function of their deadline. Delaying packets allows the network to avoid hot spotting while maintaining deadline-faithfulness. We compare RTS with another prioritizing and scheduling algorithm for real-time data dissemination in sensor networks, velocity monotonic scheduling. This paper simulates RTS based on two typical routing protocols, shortest path routing and greedy forwarding with J-Sim.

## 1. Introduction

Wireless sensor networks are an important emerging technology that will revolutionize sensing for a wide range of scientific, military, industrial and civilian applications [1]. A large number of inexpensive sensors collaborating on sensing phenomena provide cost effect detailed monitoring of the area under observation [2, 3]. While some sensor networks are deployed to collect information for later analysis [4], most applications require monitoring or tracking of phenomena in real-time [5]. Many applications require the sensor network to respond within real-time constraints. Due to the limited storage at the sensor nodes, real-time data traffic may be the main traffic in the sensor network. The solutions need to disseminate the real-time data traffic efficiently. The primary challenges here are how to prioritize and schedule packets. Moreover, due to the nature of the shared wireless medium, routing essentially provides the chance to spatially schedule the packets to reduce contention for the network resources.

## 2. Limitations of Existing Solutions

A primary challenge in real-time sensor network applications is how to carry out sensor data dissemination given source-to-sink end-to-end deadlines when the communication resources are scarce. Although routing/data transport solutions have been proposed in the context of wireless ad hoc networks, the characteristics of sensor networks make the problem different. The traffic patterns in sensor networks in response to queries or events are different from the point-to-point communication typical of sensor networks. Moreover, the bursty nature of traffic in sensor networks, as the degree of observed activity varies, can cause the network resources to be exceeded. In addition, the ad hoc nature of multi-hop sensor networks makes it difficult to schedule network traffic centrally as in traditional real-time applications.

One of the proposed solutions for real-time data dissemination [6] prioritizes packet transmission at the MAC layer according to the deadline and distance from the sink. This work has several limitations: (1) While packets are prioritized, they are not delayed when traffic is bursty, high contention results, increasing transmission and queuing delays. Furthermore, packets generated by different sensors at the same time (e.g., in response to a detected event), can lead to high collision rates. Jittering such packets can help reduce this hot-spotting; (2) MAC level solutions cannot account for the queuing delay in the routing layer (which occurs above the MAC layer); these delays can have a significant impact on end-to-end delay especially under high load; and (3) MAC level solutions require reengineering of the sensor radio hardware and firmware, making deployment difficult and potentially causing interoperability problems with earlier hardware that supports different MAC protocols. Since the scheduling needs to consider the queuing delay in the routing layer which is above the MAC layer, the impact of the routing protocols used must be carefully examined. The effect of the routing protocol on the real-time scheduling success is not sufficiently understood. Some existing solutions [7][8][9] for routing in real-time traffic context provide non-deterministic routing as an extension of stateless geographic-based routing protocols. More specifically, these approaches use the best next hop with respect to the traffic/congestion situations, not only the geographic proximity as per the greedy Geographical Forwarding protocol. In addition, Geographical Forwarding, which is used in these solutions, does not always lead to the shortest delay paths, making it more difficult to meet the deadline. Furthermore, when using a longer path in terms of number of hops, increased contention for the medium results as more transmissions are needed to forward a packet.

## 3. RTS with Virtual nodes for self stabilization: Basic Algorithms

The first distinguishing feature of RTS is that it considers all components of delay, including queuing delay at each forwarding node. The proposed scheme takes care of on demand routing along with a new concept of virtual nodes with power factor. In addition, RTS delays data packet transmission during forwarding for a duration that correlates with their remaining deadline and distance to the destination. Intuitively, this helps in heavy-traffic communication environment by making sure that priority inversion does not occur due to a node with only low priority packets sending and preventing a node with high priority packets from doing so. The virtual nodes help in reconstruction phase in fast selection of new routes. Selection of virtual nodes is made upon availability of nodes and battery status. Each route table has an entry for number of virtual nodes attached to it and their battery status. The algorithm [11] has been divided into three phases. Route Request (RReq), Route Repair (RRpr) and Error Phase (Err). Moreover, delaying the data packets before reaching the sink also helps the data aggregation/fusion and therefore energy efficiency; we do not explore this effect in this paper. Before a data packet reaches the sink, the end-to-end transmission and processing delay cannot be obtained. Therefore, we use previous measurements of delay to estimate the overall delay; we call this estimate the End-to-End Estimate of Transmission Delay (EETD) [10]. The one hop estimate is called ETD. Summing the ETD's of a data packet hop by hop during its forwarding can lead to inaccurate estimates since one hop ETD can fluctuate significantly. Therefore, we use the following function to decide the EETD:

$$EETD = ETD \cdot \frac{E2E\,Distance}{OHD} \quad (1)$$

Where OHD is One Hop Distance and the distance can be measured in different ways.

Different RTS scheduling policies can be developed based on the allocation of the available slack time among the different hops. The target transmission times are either set by the source or computed at intermediate hops based on a known algorithm. In the base RTS algorithm, the target transmission time is set to be equal at all hops and is determined as follows:

$$TD = \frac{DL - EETD}{Distance\,(X, Sink)} \cdot \alpha \quad (2)$$

Where TD be the transmission delay, DL be the deadline and the α is a constant "safety" factor for insurance that the real-time deadline would be met. For example, setting to be 0.7, would target delaying the packet 70% of the available slack time, leaving the remaining time as a safety margin.

As we can see, the Target Delay of any in-queue packet determines its priority. The time a packet is delayed in the queue can be used as the key to a priority queue that holds the packets to be transmitted. The end-to-end transmission and processing delay is considered along with the queuing delay, by taking into account the end-to-end deadline, distance and EETD.

We consider static vs. dynamic versions of the protocols depending on whether the target transmission times are set by the source and followed by intermediate nodes (static), or whether they are computed/ adjusted at intermediate nodes (dynamic).

**3.1 Static Real Time Scheduling (SRTS):** In static RTS, the target delay is set with the values of parameters at the data source. In the equation 2, the end-to-end deadline is fixed at the data source; the EETD is measured with the ETD of forwarding node and the distance from source to sink (X is the data source). So even we call it static, the different ETD's of forwarding nodes would make the target delay at each node different.

**3.2 Dynamic Real Time Scheduling (DRTS):** In dynamic RTS, the target delay is reset at each forwarding node with the local value of parameters. In equation 2, the end-to-end deadline of a packet at some forwarding node is the remaining slack time, measured by E2E Deadline −Elapsed Time. The EETD is decided by the one-hop ETD of the forwarding node and the distance from it to the sink, not the distance from source to sink. So the dynamic RTS is able to continuously refine the priority of the packet.

**3.3 Non-linear Real Time Scheduling (NLRTS):** It is also possible to allocate the available slack time non-uniformly among the intermediate hops along the path to the sink. For example, we may desire to provide the packets with additional time as it gets closer to the sink. The intuition is that in a gathering application, the contention is higher as the packet moves closer to the sink. Different policies can be developed to break down the available time. We explore the following policy:

$$TD = \frac{E2E\ deadline - EETD}{2^{\frac{RD}{OHD}}} \cdot \alpha$$

(3)

Where RD is remaining distance and OHD is one hop distance.

More generally, we may want to allocate the slack time proportionately to the degree of contention along the path. Such a heuristic may be developed by passing the contention information along with the routing advertisement and allocating the available slack time accordingly. Finally, one may decide to favor aggregation by delaying packets closer to the source where the data is more correlated.

## 4. RTS with Virtual Nodes Implementation

RTS does not ignore the queuing delay. It considers both the transmission delay and the queuing delay by doing a set of very simple scheduling decisions. The basic RTS scheduling algorithm has been shown in section 3. But RTS is more than that. Although the term RTS stands for Real-Time Scheduling, it is not only a scheduling algorithm. It involves the architecture design of the whole system. The typical architecture of a system that RTS works on is shown in figure 1. The RTS scheduler resides above (or within) the routing layer. It uses routing level information such as the end-to-end distance in making its scheduling decisions. For any real-time applications based on sensor networks, the end-to-end real-time deadline is assumed to be included on the data packet itself. Figure 1 shows an example of how this information is collected. While, in this figure, the MAC layer is shown, the RTS scheduler and the MAC layer protocol are not aware of each other.

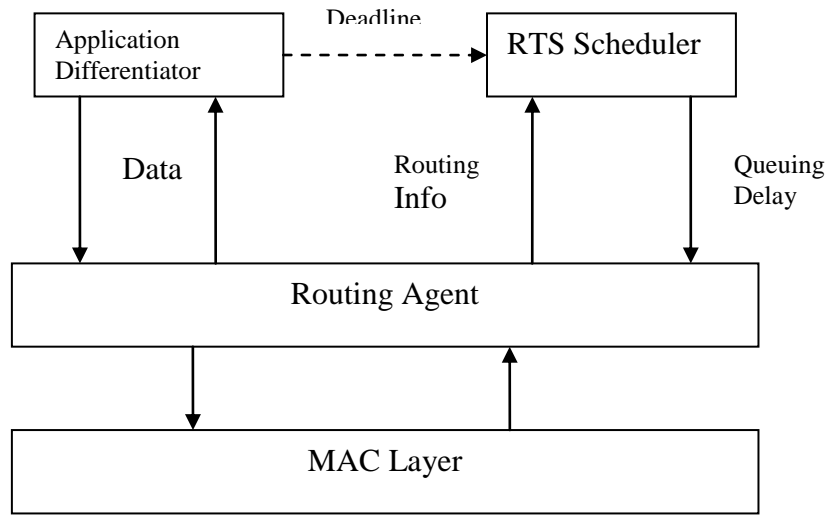

**Fig 1**: RTS Architecture

The scheme of virtual nodes has been explained with the help of an example shown in Figure 2. Assume that the node A is the source while destination is the node D. Note that the route discovered using new scheme routing protocol may not necessarily be the shortest route between a source destination pair. If the node C is having power status in critical or danger zone, then though the shortest path is A-B-C-D but the more stable path A-B-H-G-F-E-D in terms of active power status is chosen. This may lead to slight delay but improves overall efficiency of the protocol by sending more packets without link break than the state when some node is unable to process route due to inadequate battery power. The process may help when some intermediate node moves out of the range and link break occurs, in that case virtual nodes take care of the process and the route is established again without much overhead.

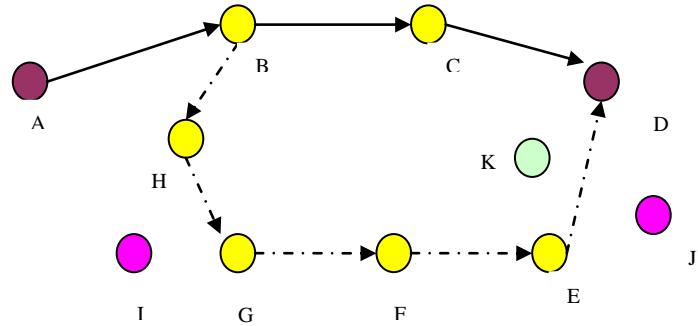

**Fig.2.** An example of routing

In Figure 2 if the node G moves out, the new establised route will be A-B-H-I-F-E-D. Here the node I is acting as virtual node (VN) for the node H and the node G. Similarly the node J can be VN for the nodes D, E, K. Virtual node (VN) has been selected at one hop distance from the said node.

## 5. RTS for different routing protocols

RTS can be adapted to work with virtually any underlying routing protocol. However, the RTS algorithm may need to be adapted to consider the cost metric used by the routing algorithm. For example, in a system based on the shortest path routing (SP), the distance parameters used by RTS scheduler is measured in number of hops. The corresponding functions are:

$$EETD = ETD \cdot E2E\ hops \qquad (4)$$

$$TD = \frac{Deadline - EETD}{H} \cdot \alpha \qquad (5)$$

$$TD = \frac{E2E\ Deadline - EETD}{2^H} \cdot \alpha \qquad (6)$$

Where H stands for end-to-end number of hops. For the geometric routing, the values of distance parameters used in RTS Scheduler would be the Euclidean distance. In summary, the following information is needed to schedule packets in RTS:

• End-to-end deadline information: this information is provided by the application in the data packet as required by any real-time data dissemination application. For those applications where the header of data packet does not include this information, an alternative way for RTS to obtain the end-to-end deadline information is needed.

• End-to-end distance information: this information is obtained from the routing protocol. For example, this information is maintained in the routing tables of traditional distance vector based or link-state based routing protocols to keep track of the cost of the path. Furthermore, in source routed protocols such as DSR, this information can be directly computed from the packet header which includes the full path to the destination. Finally, in geographic routing, Euclidian distance measured as the distance from the current node to the destination can be used as the distance metric. The output of RTS scheduler is the queuing delay, which is used by the routing protocol to decide how long to delay an incoming data packet before attempting to forward it (by passing it to the MAC layer). MAC layer prioritization is not needed by the RTS design since the packets are sent when their real time local deadline is reached; they should all be of roughly equal priority. Not requiring changes to the MAC layer is a desirable feature of RTS relative to RAP.

## 5.1 Properties of RTS

In summary, the following are the design features of the RTS framework:
- Ability to interoperate with different routing protocols: unlike the SPEED [7] or RAP [6] framework which are specific to geographical routing, RTS is not limited to a specific routing protocol. Instead, it can operate directly with any hop-based cost metric protocol and can be easily adapted to work with Geometric routing protocols. This flexibility is demonstrated via simulation later in this paper.

- This scheme utilizes a mesh structure and alternate paths in case of failure. The scheme can be incorporated into any on-demand unicast routing protocol to improve reliable packet delivery in the face of node movements and route breaks. Alternate routes are utilized only when data packets cannot be delivered through the primary route. As a case study, the proposed scheme has been applied to QDPRA [11] and it was observed that the performance improved. Simulation results indicated that the technique provides robustness to mobility and enhances protocol performance. It was found that overhead in this protocol was slightly higher than others, which is due to the reason that it requires more calculation initially for checking virtual nodes.

- Soft Real-time: RTS maintains a uniform delivery speed of data packets, meeting the deadline of most data traffic with best effort. Packets that pass their deadline are not dropped. While it's possible to better support hard real-time in this framework (for example, by increasing the safety margin, and immediately dropping packets that are late), we do not pursue such extensions.

- No MAC layer support required: Unlike the SPEED or RAP, RTS does not require MAC layer support for prioritized scheduling (as with RAP) or for tracking delay (as with SPEED). This makes RTS readily deployable on existing infrastructure.

- QoS routing: RTS integrates the transmission delay with the queuing delay, considering both the lower layer communication cost and that of higher layers and differentiating the data flows with different real-time constraints.

- Ability to withstand high load and hot spotting: RTS uses the queuing mechanism to delay any data flows to restrict contention to occur among only the most urgent traffic. This allows RTS to gracefully accommodate higher traffic levels than RAP or SPEED.

- Data Fusion: RTS tries to delay any incoming data traffic which gives more possibility of the data aggregation operations. Since the data aggregation is a primary data operation during the data forwarding for most applications, RTS fits better than the other approaches which attempt to send packets without delay.

## 6. Implementation and Experimental Evaluation

We implemented RTS (Static, Dynamic and Non-Linear) with both the Shortest Path routing and Greedy Geographic Forwarding in the Network Simulator (J-SIM 2). We also implemented the RAP Velocity Monotonic Scheduling (VMS) with GF, including the specialized MAC support required by it on J-SIM per the specification. Since GF has been shown to significantly outperform traditional routing protocols such as DSR [10] and deadline-based scheduling, in the context of sensor network data dissemination, we restrict the routing comparison to GF and SP, and the scheduling comparison to VMS(Velocity Monotonic Scheduling)and RTS.

**Table 1:** Simulation Parameters

| Mac layer protocol | IEEE 802.11 with prioritizing extension |
|---|---|
| Transmission Radio Range | 250 m |
| Bandwidth | 2 Mbps |
| Data Packet Size | 32 B |
| Data Rate | 2 packets/second |
| Simulation Area | $1000 \times 1000$ m2 |
| Number of Sensor nodes | 100 |
| Effective Simulation Time | 120 sec |

Table1 shows the simulation parameters we use; unless otherwise indicated these parameters are used in the studies. We use both the grid and random deployment to simulate our algorithm. In grid deployment, we divide the covered simulation area into a $10 \times 10$ grid. One of the 100 sensor nodes is placed at the center of each the grid tiles. The sink is placed on the northwest corner of the network. Nodes publish data at the rate of 2 packets per second in order to simulate a fairly high load traffic scenario. In random deployment, the 100 nodes are randomly placed in the simulation area while the sink is placed roughly at the center of the area. First, we compared RTS with VMS both using the same routing protocol (GF); recall that GF was used in the original RAP scheme [6]. Later, we show that SP significantly outperforms GF for RTS. Since we consider soft real-time applications, a change we made to the RAP mechanism is that each node tries to forward all incoming data packets, no matter if the deadline is already missed or not. In the original implementation of RAP, the packets missing the deadline would be dropped. Since RTS does not require any MAC layer information, we use the original IEEE 802.11 as our MAC layer protocol. We considered the issue of what the RTS safety margin parameter α should be set to. If α is too high, packet delay variability can cause deadlines to be missed since most of the slack time is taken up by intentional RTS delay and unexpected delays cause a packet to miss the deadline. Conversely, if α is too low, packets are conservatively sent quickly towards the sink, possibly overflowing buffers around it. Experimentally, we observed that a safety margin parameter of 0.7 works well across different deadlines. Thus, 30% of the deadline budget is set aside to account for inaccuracies in ETD estimates and/or unexpected transmission or queuing delays.

The first experiment studies the performance of RTS scheduling for sensor networks relative to RAP. Figure 3 shows that for different packet requirement, the miss ratios and drop ratios of RTS Static and Dynamic are much lower than those of DVM and SVM for across all the considered deadline range. Dynamic RTS outperforms static RTS in terms of the miss ratio.

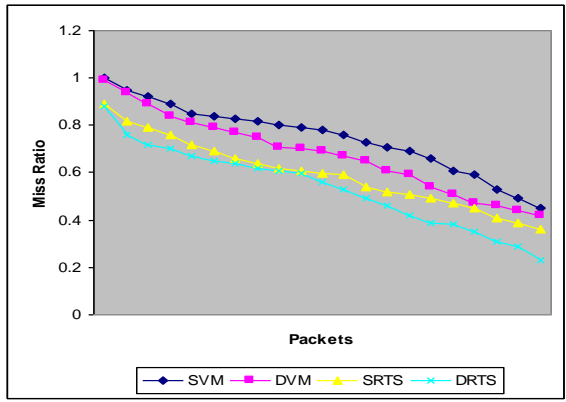
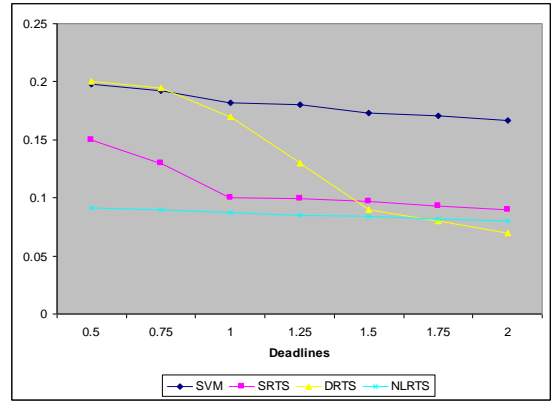

Figure 3 Miss Ratio                    Figure 4 Deadline

## 6.1 Performance under Random Deployment

RTS and VMS were also evaluated using a random deployment scenario where the 100 nodes were randomly placed within the simulation area. Three random deployments of 100 sensor nodes in a 1000 × 1000m2 areas are taken. Each result represents the average of several experiments with different seeds. We varied the deadline requirements from 0.5 to 2.0 seconds in steps of 0.5 seconds. Ratios and drop ratios for the different algorithms. The simulations show that both RTS and VMS (figure 4) perform much better in random scenarios than they did in the grid scenarios possibly because the location of the sink is central to the simulation area, making the average sensor distance to the sink smaller. Again RTS provides superior performance to VMS. For the VMS, the drop ratios do not decrease as the deadline grows since it prioritizes but does not delay packets. The drop ratio becomes the lower bound of the miss ratio. RTS shows more reactivity since both the drop ratio and miss ratio keep decreasing as the deadline requirement is relaxed.

**6.2 Performance under Busty Traffic**
In this study, we evaluate the performance of RTS vs. RAP under busty traffic conditions. Each node publishes alternately publishes packets at the pre-set data rate for 5 Seconds then stops publishing for the second 5. Figure 5 shows the miss ratios and drop ratio of RTS and SVM under this busty traffic with end-to-end deadline from 0.1 second to 3.0 seconds. From the figure we can see that the miss ratio of dynamic RTS is much lower than that of SVM with the busty traffic, because RTS can tolerate the traffic burst by delaying some packets, and taking advantage of the idle period. On the other hand, SVM cannot make use of the traffic behavior since it does not delay packets. The decrease in the drop ratio shows that RTS also deliveries more packets as the deadline constraints are relaxed.

**6.3 Comparison with SPEED**
We also built simulation models for the SPEED framework [7] within the Java simulator. Unfortunately, the simulation results we obtain do not match the performance demonstrated in the original SPEED papers [7][8].

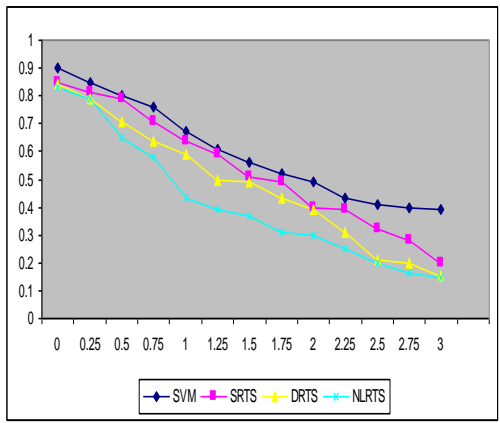

Figure 5 Busty traffic

We implemented the full specification of SPEED, SPEED-T (Minimal one hop delay first), and SPEED-S (maxim alone hop progress speed first), simulating them in the exactly same scenarios specified in [7]. Our experiences with SPEED show that it performs extremely poorly at high loads because its backpressure mechanism is not suited to the situations where alternative paths are also congested. In those situations backpressure ends up increasing the load on the network by routing packets through unnecessarily long paths. We believe that our comparison is fair because all the algorithms are implemented in the same environment (thus removing any differences that occur due to the different simulators). Because of the overall poor performance under high load, we do not compare RTS with SPEED in detail.

## 7. Conclusions & Future Work

Real-time data dissemination is a service of great interest to many sensor network applications. The paper proposed and evaluated the real time scheduling mechanism for real-time sensor network applications. RTS offers significant advantages over existing real-time sensor data dissemination schemes. It accomplishes real-time support by delaying packets a fraction of their slack time at each hop. As a result, it is better able to tolerate busts than schemes that simply prioritize packet transmission.

RTS can operate with simple routing protocols easily and outperforms RAP in both the miss ratio and overall delay. The paper explored criteria for allocating the available slack time among the different nodes and showed that nonlinear distribution of the slack time, with more time assessed to hops closer to the sink results in better performance than linear distribution of the slack time in the gathering scenarios that we studied. RTS is a network layer solution and does not require changes to lower level protocols making it easier to deploy and independent of the underlying sensor network hardware capabilities. Using simulation, we found the drop ratio is the lower bound of the miss ratio of real-time communication. If the drop ratio is decreased, given a reasonable end-to-end deadline, the miss ratio of these real-time applications should also be decreased. Mostly the packets are dropped due to congestion as the network capacity is exceeded.